\documentstyle[prb,aps,floats,epsfig]{revtex}

\begin{document}
\twocolumn[\hsize\textwidth\columnwidth\hsize\csname @twocolumnfalse\endcsname

\title{Density-functional theory and the {\it v}-representability
problem for model strongly correlated electron systems}
\author{Arno Schindlmayr and R. W. Godby}
\address{Cavendish Laboratory, University of Cambridge, Madingley
Road,\\ Cambridge CB3 0HE, United Kingdom}
\date{Received 28 November 1994}
\maketitle

\begin{abstract}
Inspired by earlier work on the band-gap problem in insulators, we
reexamine the treatment of strongly correlated Hubbard-type models
within density-functional theory. In contrast to previous studies,
the density is fully parametrized by occupation numbers {\em and\/}
overlap of orbitals centered at neighboring atomic sites, as is the
local potential by the hopping matrix. This corresponds to a good
formal agreement between density-functional theory in real space and
second quantization. It is shown that
density-functional theory is formally applicable to such systems and
the theoretical framework is provided. The question of
noninteracting {\it v} representability is studied numerically for
finite one-dimensional clusters, for which exact results are
available, and qualitatively for infinite systems. This leads to the
conclusion that the electron density corresponding to interacting
systems of the type studied here is in fact {\em not} noninteracting
{\it v} representable because the Kohn-Sham electrons are unable to
reproduce the correlation-induced localization correctly.
\end{abstract}
]
\narrowtext

\section{Introduction} \label{Introduction}

Originally formulated in the 1960s, density-functional theory (DFT)
has since become the most widely used method for electronic-structure
and total-energy calculations. It is a general theory that is
applicable to finite molecular systems as well as to bulk solids and,
unlike other computational methods, it has the distinct advantage of
being {\em in principle\/} an exact theory that takes the electronic
interaction fully into account. Density-functional theory is based on
the observation that the total energy and other characteristic
ground-state properties of a system of interacting electrons in an
external potential can be considered as unique functionals
of the one-particle density $n({\bf r})$, which then
replaces the wave functions as the basic variable.

Although potentially powerful, the original Hohen\-berg-Kohn theorem
\cite{Hohenberg-Kohn} does not provide a simple recipe to calculate
the ground-state density of the interacting electron system, so that
practical applications of density-functional theory rely on the
Kohn-Sham scheme,\cite{Kohn-Sham} in which the interacting system is
replaced by a fictitious system of noninteracting electrons with the
same spatial density moving in an effective potential. The central
{\em assumption\/} is that such an equivalent system of
noninteracting electrons always exists, i.~e., that all interacting
{\it v} representable densities are also noninteracting {\it v}
representable.

In the Kohn-Sham scheme, the total energy is split into several
contributing functionals, all but one of which are either known
explicitly or can be replaced by other well-known terms. The essential
unknown quantity is the exchange-correlation energy $E_{\rm xc}[n]$, a
universal functional of the density, and its functional derivative,
the exchange-correlation potential $V_{\rm xc}([n];{\bf r})$.

Although, unfortunately, the exact analytic expression for the
exchange-correlation energy remains unknown, many properties of the
functional have been cataloged.\cite{Levy-Perdew} One of the more
surprising features is a finite discontinuity in the exchange-correlation
potential with respect to particle number,\cite{band-gap} which has the
effect that even exact density-functional theory will underestimate the
band gap of semiconductors. The question whether the poor agreement
between density-functional calculations and the true band gap is indeed
mainly due to this discontinuity or rather to additional approximations
such as the local-density approximation (LDA) is known in the literature
as the band-gap problem.

The Hubbard model, a simple second-quantization model for strongly
correlated electrons, may be solved exactly for small clusters and
so could be used to explore the properties of exact density-functional
theory. This was done by Gunnarsson and Sch\"onhammer in a study of the
band-gap problem \cite{GS-band-gap} and subsequently for comparing the
Kohn-Sham Fermi surface with the exact quasiparticle Fermi
surface.\cite{GS-Fermi-surface} In their second-quantization version of
density-functional theory, the density is replaced by the occupation
numbers only and the place of the external potential is taken by the
diagonal on-site energies, whereas all nondiagonal hopping parameters are
treated as constants. This is a consistent approach to correlated systems,
but it does not follow the second quantization of the operators in
question.

In this paper, we investigate a different formulation of
density-functional theory for correlated Hubbard-type systems that is
closer to the original scheme in the continuum. Our approach is
strictly based on second quantization with a complete
parametrization of the density, which includes the overlap between
orbitals centered at different atomic sites, and no constraints on the
local potential. We demonstrate that density-functional theory,
in this formulation, is {\em in principle\/} applicable to
the Hubbard model, but we also show that the ground-state density
is in fact {\em not\/} noninteracting {\it v} representable under
the Kohn-Sham scheme. These findings should be of particular interest
since the results of the above-mentioned model calculations have often
been cited in connection with the properties of real-space
functionals, especially the band-gap problem:\cite{Dreizler-Gross}
Gunnarsson and Sch\"onhammer found that the discrepancy between the
true band gap and the energy-eigenvalue gap derived by
density-functional theory using the LDA was mainly due to the LDA
itself rather than to the discontinuity in the exchange-correlation
potential; the opposite was found in a study of more realistic
semiconductor models by Godby, Schl\"uter and Sham.\cite{Godby}

In Section \ref{Formalism} we will establish a formulation of
density-functional theory appropriate for the Hubbard model and prove
the fundamental theorems. In Section \ref{Correlation} we will
examine the influence of correlation and the external potential on
the density and highlight the importance of overlap between orbitals
centered at different sites. In Section \ref{Results}
we will discuss the question of {\it v} representability and
demonstrate that the ground-state density of the interacting Hubbard
model and of related models is {\em not\/} noninteracting
{\it v} representable.

\section{Formalism} \label{Formalism}

\subsection{The Hubbard model}

The Hubbard Hamiltonian \cite{Hubbard} is given by
\begin{equation} \label{Hamiltonian3D}
\hat{H} = \sum_{\langle i j \rangle} \sum_\sigma t_{ij} c^\dagger_{i\sigma}
c_{j\sigma} + U \sum_i \hat{n}_{i\uparrow} \hat{n}_{i\downarrow} .
\end{equation}
The index $i$ labels the atomic sites and the summation
over $\langle i j \rangle$ includes all pairs up to nearest neighbors.
$\sigma \in \{\uparrow,\downarrow\}$ denotes the two spin
orientations.

The first part of the Hamiltonian represents the contribution due to
the kinetic energy and the external (pseudo)potential $V_{\rm ext}({\bf
r})$ and is responsible for the electrons hopping between the atomic
sites. $c^\dagger_{i\sigma}$ and $c_{i\sigma}$ are the creation
and annihilation operators for an electron at site $i$ with spin
$\sigma$, respectively. The hopping-matrix elements $t_{ij}$ are real
and satisfy the symmetry condition $t_{ij}=t_{ji}$. They are the
representation of the external potential in second quantization
and are related to it by
\begin{equation} \label{t-def}
t_{ij} = \int\! \phi^*({\bf r}-{\bf R}_i) \left(
-\frac{\hbar^2}{2m_e} \nabla^2 + V_{\rm ext}({\bf r}) \right)
\phi({\bf r}-{\bf R}_j) \, d{\bf r} ,
\end{equation}
where $\phi({\bf r}-{\bf R}_i)$ is the Wannier-type orbital
centered at ${\bf R}_i$ of the system in question. The existence of
these orbitals underlies second quantization,
but their actual analytic form is of no consequence for the numerical
work presented below.

The second contribution to the Hamiltonian represents the on-site
Coulomb repulsion, which acts between two electrons occupying the
same atomic site. $\hat{n}_{i\sigma} \equiv c^\dagger_{i\sigma}
c_{i\sigma}$ is the particle-number operator for site
$i$ and spin $\sigma$. The interaction parameter $U$ is real and
$U>0$. Its value is given by
\begin{equation} \label{U-def}
U = \frac{1}{2} \int \!\!\! \int \!
\frac{e^2}{4\pi\varepsilon_0}\, \frac{ |\phi({\bf r})|^2 \,
|\phi({\bf r}')|^2 }{|{\bf r}-{\bf r}'|}\, d{\bf r}\,
d{\bf r'} .
\end{equation}

When discussing the adjusted density-functional formalism in this
section, we will always refer to the Hubbard Hamiltonian in the
general form (\ref{Hamiltonian3D}) without specifying the
dimensionality or configuration of the system. In particular, no
spatial symmetries are assumed, so the formalism holds for
lattices as well as for finite or disordered systems. In the latter
case, the occurence of different Wannier orbitals may lead to a
site-dependent interaction parameter, but this is not in conflict
with the general formalism.
$M$ denotes the number of sites and $N$ is the number of electrons.

As a prerequisite of density-functional theory, it is essential to
define the density in a proper way. As for the Hamiltonian, we assumed
that overlap of the Wannier orbitals was negligible except between
neighboring sites, and this assumption must be retained for
consistency.
The density corresponding to some state $|\Psi\rangle$ will therefore
be of the general form
\begin{equation} \label{density}
n({\bf r}) = \langle \Psi | \hat{n}({\bf r}) | \Psi
\rangle = \sum_{\langle i j \rangle} n_{ij} \phi^*({\bf r}-{\bf
R}_i) \phi({\bf r}-{\bf R}_j) .
\end{equation}
The density operator $\hat{n}({\bf r})$ is given by
\begin{equation}
\hat{n}({\bf r}) = \sum_{\langle ij \rangle} \sum_\sigma
\phi^*({\bf r}-{\bf R}_i) \phi({\bf r}-{\bf R}_j)
c^\dagger_{i\sigma} c_{j\sigma}
\end{equation}
in second quantization and the density is thus para\-me\-trized by the
coefficients
\begin{equation}
n_{ij} = \sum_\sigma \langle \Psi | c^\dagger_{i\sigma}
c_{j\sigma} | \Psi \rangle .
\end{equation}
The diagonal elements $n_i \equiv n_{ii}$ are just the
orbital occupation numbers, but we are also left
with off-diagonal elements. Strictly speaking, the dependence of the
total energy on the density coefficients is that of a function rather
than a functional, but for clarity, we will continue talking
about density functionals. It should then be borne in mind that all
densities to be considered are of the form (\ref{density}).

\subsection{The Hohenberg-Kohn theorem}

The analog of the Hohenberg-Kohn theorem in this formalism
incorporates three important statements.

(i) The ground-state expectation value of any
observable $\hat{O}$ is a unique functional $O[n^{\rm GS}]$ of
the ground-state electron density $n^{\rm GS}$.

(ii) The ground-state density $n^{\rm GS}$ minimizes the total-energy
functional $E[n]$.

(iii) The total-energy functional can be written in the form
\begin{equation} \label{int-energy}
E[n] = F[n] + \sum_{\langle ij \rangle} t_{ij} n_{ij}
\end{equation}
where $F[n]$ is a {\em universal\/} functional of the density.

In proving these statements, we have largely followed the procedure
given by Levy.\cite{Levy} Throughout, the ground
state is assumed to be nondegenerate, but the formalism can be
extended to cover degenerate ground states in much the same way as
conventional density-functional theory.\cite{Kohn}
First, we define the energy functional by
\begin{equation} \label{Efunctional}
E[n] := \min_{|\Psi\rangle \to n} \langle \Psi | \hat{H}
| \Psi \rangle .
\end{equation}
The constraint $|\Psi\rangle \to n$ marks all antisymmetric
$N$-body wave functions that yield the
density $n$ as defined above. This functional is properly defined for
all $N$-representable densities, i.~e., all densities that can be
constructed from $N$-fermion wave functions. In particular, this
includes all ground-state densities studied here. The variational
principle then guarantees that the ground-state density indeed
minimizes $E[n]$.

At the same time, the definition (\ref{Efunctional}) constitutes a map
$n^{\rm GS} \mapsto |\Psi[n^{\rm GS}]\rangle$, where $|\Psi[n^{\rm GS}]\rangle$
denotes the $N$-body wave function that yields the ground-state
density $n^{\rm GS}$ of a particular system and minimizes the
energy. We can therefore properly define the ground-state expectation
value of an arbitrary observable $\hat{O}$ as a functional of the
ground-state density by
\begin{equation}
O[n^{\rm GS}] := \langle \Psi[n^{\rm GS}] | \hat{O} | \Psi[n^{\rm GS}]
\rangle .
\end{equation}

The single-particle contribution to the functional $E[n]$, which
we will henceforth term the hopping energy, can be essentially
simplified. Using the explicit form of the Hubbard Hamiltonian and
the definition of the density coefficients, (\ref{Efunctional})
can be transformed into the expression
\begin{equation}
E[n] = \sum_{\langle ij \rangle} t_{ij} n_{ij} +
\min_{| \Psi \rangle \to n} \left\langle \Psi \left| U \sum_i
\hat{n}_{i\uparrow} \hat{n}_{i\downarrow} \right| \Psi \right\rangle .
\end{equation}
We can now define the observable $\hat{F}$ by
\begin{equation}
\hat{F} := U \sum_i \hat{n}_{i\uparrow} \hat{n}_{i\downarrow}
\end{equation}
and thereby obtain the form (\ref{int-energy}).
$F[n]$ is universal in the sense that it is independent of
the particle number and the external potential. The upper
limit in the sum over $i$ need not be specified because
the number of sites enters through the constrained
search over the wave functions that yield the density $n$.
Finally, the interaction $U$ is fixed through its
definition (\ref{U-def}). This completes the proof of
the Hohenberg-Kohn theorem.

\subsection{The Kohn-Sham scheme}

The second part of the Hohenberg-Kohn theorem suggests that if we
knew the exact analytic form of $E[n]$, we could find the
ground-state energy by a variational principle that minimizes the
functional under the constraint of particle conservation, but it does
not provide a practical procedure to do so. Rather, the constrained
search over the wave functions leaves us with another complicated
$N$-body theory.

The essential simplification is obtained by the Kohn-Sham scheme,
which {\em assumes\/} that the ground-state density $n^{\rm GS}$ of the
Hamiltonian (\ref{Hamiltonian3D}) can alternatively be generated by
the ground state of a fictitious system of noninteracting electrons
moving in an effective potential. This is known as the assumption of
noninteracting {\it v} representability. The equivalent system
can be used to calculate the exact ground-state energy and the
mathematical expense is reduced to an effective one-particle
problem. This one-particle Hamiltonian is given by
\begin{equation} \label{effHamiltonian3D}
\hat{H}_s = \sum_{\langle ij \rangle} \sum_\sigma t^{\rm eff}_{ij}
c^\dagger_{i\sigma} c_{j\sigma} .
\end{equation}
The $N$-body ground state is supposed to be
nondegene\-rate. Following the central assumption stated above,
the hopping parameters are chosen so that the ground-state density
matches that of the interacting system. In this case, the
density coefficients are given by
\begin{equation}
n^{\rm GS}_{ij} = \sum_{\gamma = 1}^N \sum_\sigma \langle
\psi_\gamma | c^\dagger_{i\sigma} c_{j\sigma} |
\psi_\gamma \rangle
\end{equation}
in terms of the one-particle wave functions that correspond
to the $N$ lowest energy eigenvalues $\epsilon_\gamma$ of $\hat{H}_s$.
The ground-state energy of the noninteracting system is a functional
of $n^{\rm GS}$ and can be written directly as
\begin{equation}
E_s[n^{\rm GS}] = \sum_{\gamma = 1}^N \epsilon_\gamma =
\sum_{\langle ij \rangle} t^{\rm eff}_{ij} n^{\rm GS}_{ij} .
\end{equation}
Due to the Hohenberg-Kohn theorem, this functional has to
be stationary under infinitesimal density variations $\delta n$
within the domain of valid $N$-fermion densities:
\begin{equation} \label{non-int-variation}
0 = \delta E_s = \sum_{\langle ij \rangle} t^{\rm eff}_{ij} \delta n_{ij} .
\end{equation}

In order to map the interacting electrons onto the noninteracting
system, we start by separating a Hartree term out of the
total-energy functional (\ref{Efunctional}):
\begin{equation} \label{E-Hartree-XC}
E[n] = \sum_{\langle ij \rangle} t_{ij} n_{ij} + \frac{U}{2} \sum_i
n_i^2 + E_{\rm xc}[n] .
\end{equation}
The exchange-correlation functional $E_{\rm xc}[n]$
incorporates all exchange and correlation effects that are not
included in the Hartree term and is formally defined by
\begin{equation} \label{XC-def}
E_{\rm xc}[n] := F[n] - \frac{U}{2} \sum_i n_i^2 .
\end{equation}
At the ground-state density, $E[n]$ must also be stationary under
infinitesimal variations $\delta n$ that correspond to $N$-fermion
densities:
\begin{equation} \label{Evariation}
0 = \delta E = \sum_{\langle ij \rangle} \left( t_{ij} + U \delta_{ij} n^{\rm GS}_i
+ v^{\rm xc}_{ij}[n^{\rm GS}] \right) \delta n_{ij} ,
\end{equation}
where $v^{\rm xc}_{ij}[n^{\rm GS}]$ denotes the elements of the
exchange-correlation matrix, which is defined by
\begin{equation} \label{XC-matrix}
v^{\rm xc}_{ij}[n^{\rm GS}] := \left. \frac{\partial
E_{\rm xc}[n]}{\partial n_{ij}} \right|_{n=n^{\rm GS}} .
\end{equation}
By comparison with (\ref{non-int-variation}), the
effective hopping-matrix elements can now be uniquely defined as
\begin{equation} \label{eff-parameters}
t^{\rm eff}_{ij} := t_{ij} + U \delta_{ij} n^{\rm GS}_i +
v^{\rm xc}_{ij}[n^{\rm GS}] .
\end{equation}
This expression for the effective potential can
be used to calculate the ground-state electron density in a
self-consistent manner. The total energy can eventually be calculated
from (\ref{E-Hartree-XC}). As usual, it is possible to replace the
term that contains the kinetic energy using the eigenvalues
$\epsilon_\gamma$ of the one-particle Schr\"odinger equation, but
this is not advisable here because the hopping term is
already of the simplest possible form.

In principle, splittings of the total energy other than
(\ref{E-Hartree-XC}) are possible. In Ref.\ \onlinecite{GS-band-gap} the
hopping term is treated much like the kinetic energy in the conventional
formalism. Hence the functional $E[n]$ splits into the hopping energy of
the corresponding system with $U=0$, the Hartree term, and an
exchange-correlation energy defined in a different way from the one used
here. However, only the formulation introduced here shows the distinct
feature that the term $F[n]$ is a {\em universal\/} functional independent
of the external potential, in practice a prerequisite for any systematic
application of density-functional theory, and a feature of
density-functional theory for real systems. We therefore believe that our
formulation of density-functional theory is the natural counterpart of
that used for {\it ab initio} calculations.

\subsection{The problem of {\it v} representability}

The conceptual idea of the Kohn-Sham scheme, constructing the
ground-state density by means of a system of noninteracting
electrons, ultimately leads to the question of {\it
v} representability. As a definition, we call a density
interacting {\it v} representable if it corresponds to the ground
state of a Hamiltonian of the form (\ref{Hamiltonian3D}), which
includes a specified on-site interaction term. On the other hand, a
density that corresponds to the ground state of a one-particle
Hamiltonian of the form (\ref{effHamiltonian3D}) is called
noninteracting {\it v} representable.

The functionals in second quantization differ from those in the
continuum formulation and we cannot hope to transfer any previous
findings on the question of {\it v} representability. The
constrained search guarantees that all functionals are well defined
for arbitrary $N$-representable densities, which includes both
interacting and noninteracting {\it v} representable densities, but
the crucial question is whether the ground-state density of a given
interacting system indeed minimizes a corresponding Kohn-Sham energy
functional $E_s[n]$. As there is no obvious indication whether the two
domains of interacting and noninteracting {\it v} representable densities
overlap or even coincide, we have investigated this question by means
of numerical simulations and we have come to the conclusion that they
are in fact largely distinct.

\section{Correlation and the Overlap Coefficients} \label{Correlation}

\subsection{The effect of correlation}

We consider a finite one-dimensional Hubbard chain at half-filling,
which we solve numerically by exact diagonalization. We assume the
on-site energies $e_i \equiv t_{ii}$ to be zero and the other
hopping parameters to be identical between all nearest neighbors:
\begin{equation} \label{Hamiltonian1D}
\hat{H} = -t \sum_{i=1}^{M-1} \sum_\sigma \left(
c^\dagger_{i\sigma} c_{(i+1)\sigma} +
c^\dagger_{(i+1)\sigma} c_{i\sigma} \right) + U
\sum_{i=1}^M \hat{n}_{i\uparrow} \hat{n}_{i\downarrow}
\end{equation}
with $t>0$. We give results for an eight-site chain here,
a configuration that has a nondegenerate ground state for all
choices of $t$ and $U$. Calculations performed for other
chain lengths have produced very similar results.

To visualize the effect of correlation, we fix $t=1$ and calculate
the ground-state energy and the density coefficients for
different $U$. The results are shown in Table \ref{E-of-U}.
The total energy $E(U)$ is the lowest eigenvalue of the
Hamiltonian matrix, $T(U) = -t \sum_{i=1}^{M-1} ( n_{i(i+1)} +
n_{(i+1)i} )$ denotes the hopping energy and $E_H(U) = U
\sum_{i=1}^M n_i^2 /2$ is the Hartree energy. The
exchange-correlation
energy is calculated from $E_{\rm xc}(U) = E(U) - T(U) - E_H(U)$.

\begin{table}
\caption{Energy and density coefficients of the eight-site Hubbard chain
for $t=1$ and selected values of $U$. The increasing correlation reduces
the overlap and thus leads to a strong localization of the electrons. In
the limit $U \to \infty$, the orbital overlap reaches zero and the
electrons are completely localized.} \label{E-of-U}
\begin{tabular}{lrrrrr}
& $U=0$ & $U=2$ & $U=4$ & $U=8$ & $U\to\infty$ \\
\hline
$E(U)$ & -9.52 & -6.23 & -4.24 & -2.42 & 0.00 \\
$T(U)$ & -9.52 & -8.82 & -7.19 & -4.59 & 0.00 \\
$E_H(U)$ & 0.00 & 8.00 & 16.00 & 32.00 & $\infty$ \\
$E_{\rm xc}(U)$ & 0.00 & -5.41 & -13.05 & -29.83 & $-\infty$ \\
\\
$n_1 = n_8$ & 1 & 1 & 1 & 1 & 1 \\
$n_2 = n_7$ & 1 & 1 & 1 & 1 & 1 \\
$n_3 = n_6$ & 1 & 1 & 1 & 1 & 1 \\
$n_4 = n_5$ & 1 & 1 & 1 & 1 & 1 \\
$n_{12} = n_{78}$ & 0.862 & 0.793 & 0.641 & 0.408 & 0.000 \\
$n_{23} = n_{67}$ & 0.495 & 0.455 & 0.370 & 0.238 & 0.000 \\
$n_{34} = n_{56}$ & 0.758 & 0.714 & 0.589 & 0.376 & 0.000 \\
$n_{45}$ & 0.529 & 0.484 & 0.391 & 0.249 & 0.000 \\
\end{tabular}
\end{table}

We find that the occupation numbers $n_i = 1$ exactly, independent of
$U$: the electrons are distributed evenly. On the other hand,
the interaction obviously reduces the overlap between orbitals
centered at different sites. The underlying physical reason is that
the on-site Coulomb repulsion opposes the hopping of electrons to a
site that is already occupied. The correlation thus reduces the
electron fluctuations. For increasing $U$, the
electrons become eventually localized on the atomic sites and
the hopping energy approaches zero.

The localization effect that we have visualized here is just the
mechanism that is responsible for the Mott metal-insulator
transition,\cite{Mott} although it has been shown that a true phase
transition does not occur for finite $U$: the half-filled Hubbard chain
is a conductor for $U=0$ and a Mott insulator otherwise.\cite{Lieb-Wu}

\subsection{The importance of the overlap parameters}

The possibility of electrons hopping between neighboring atomic sites
requires a significant overlap of the orbitals. On the
mathematical side, this overlap explicitly enters the expression for
the hopping parameters (\ref{t-def}). It is therefore both
qualitatively and quantitatively an important feature of the system.
For reasons of consistency, one must use overlap coefficients as
well as occupation numbers to parametrize the density. On the other
hand, it might be argued that the overlap
contribution to the density in real space was negligible due to the
strong localization of the Wannier-type orbitals. Although this
argument might seem plausible if the Wannier states are chosen to
resemble the localized {\it d} orbitals found in transition metals,
there is more to be said: the energy as a functional of the density
depends very sensitively on the overlap coefficients and it is
therefore essential to include those in all practical applications.
To clarify this point, we now hold the
parameter $U=4$ fixed and vary $t$. Results are given in Table
\ref{E-of-t}. The density coefficients are the same as in Table
\ref{E-of-U}, because they depend on the ratio $t/U$ only.

As the interaction is now specified by fixing $U$, all energy
expectation values are unique functionals of the density due to the
Hohenberg-Kohn theorem. The important point to note is that for this
particular configuration, the occupation numbers are identical
regardless of the external potential represented by $t$, {\em but the
overlap coefficients are not}, which allows us to identify their
influence. As for the hopping energy, the dependence
on the overlap is known explicitly:
\begin{equation}
T[n] = -t \sum_{i=1}^{M-1} \left( n_{i(i+1)} + n_{(i+1)i} \right) .
\end{equation}
The exchange-correlation energy $E_{\rm xc}[n]$ also varies,
but its dependence on the overlap is much more subtle. When applying
the Kohn-Sham scheme, whether in exact or approximate
density-functional theory, we must thus make sure to reproduce both
the exact occupation numbers {\em and\/} overlap coefficients of the
interacting system.

In the formalism used in Ref.\ \onlinecite{GS-band-gap}, $T[n]$ is equal
to the hopping energy of the corresponding system with zero interaction
and would therefore be proportional to $t$, whereas the Hartree term is
constant if the occupation numbers are. The sum of both then changes
linearly with $t$. However, from the values given in Table \ref{E-of-t},
it is clear that the total energy is not linear. The exchange-correlation
energy must therefore depend sensitively on $t$, which is treated as a
parameter. If an analytic approximation is used rather than the exact
functional, it should definitely show this feature. It is then clear that
an approximation such as the ``local-density approximation'' defined by
Eq.\ (2) in Ref.\ \onlinecite{GS-band-gap}, which depends only on $U$ and
on the occupation numbers and so is independent of $t$, has a fundamental
flaw not shared by the normal LDA for real systems. Results that depend
on making a connection between such an approximation and the LDA for real
systems, such as the conclusions of Ref.\ \onlinecite{GS-band-gap}
regarding the LDA band-gap error, are therefore questionable.

\vfill
\begin{table}[h]
\caption{Energy and density coefficients of the eight-site Hubbard chain
for $U=4$ and selected values of $t$. As the occupation numbers remain
constant, the change in the energy values reflects the dependence of the
functionals on the nondiagonal density coefficients.} \label{E-of-t}
\begin{tabular}{lrrrrr}
& $t\to\infty$ & $t=2$ & $t=1$ & $t=0.5$ & $t=0$ \\
\hline
$E[n]$ & $-\infty$ & -12.45 & -4.24 & -1.21 & 0.00 \\
$T[n]$ & $-\infty$ & -17.64 & -7.19 & -2.29 & 0.00 \\
$E_H[n]$ & 16.00 & 16.00 & 16.00 & 16.00 & 16.00 \\
$E_{\rm xc}[n]$ & -8.00 & -10.81 & -13.05 & -14.92 & -16.00 \\
\\
$n_1 = n_8$ & 1 & 1 & 1 & 1 & 1 \\
$n_2 = n_7$ & 1 & 1 & 1 & 1 & 1 \\
$n_3 = n_6$ & 1 & 1 & 1 & 1 & 1 \\
$n_4 = n_5$ & 1 & 1 & 1 & 1 & 1 \\
$n_{12} = n_{78}$ & 0.862 & 0.793 & 0.641 & 0.408 & 0.000 \\
$n_{23} = n_{67}$ & 0.495 & 0.455 & 0.370 & 0.238 & 0.000 \\
$n_{34} = n_{56}$ & 0.758 & 0.714 & 0.589 & 0.376 & 0.000 \\
$n_{45}$ & 0.529 & 0.484 & 0.391 & 0.249 & 0.000 \\
\end{tabular}
\end{table}
\vspace{-3ex}

\section{Results and Discussion} \label{Results}

\subsection{Numerical studies for small $M$}

In order to test the existence of effective hopping parameters and
thereby decide the question of noninteracting {\it
v} representability, we set up a trial Hamiltonian matrix and use
iterative nonlinear optimization techniques to vary the effective
on-site energies and nearest-neighbor hopping
parameters so as to reproduce as closely as possible the density
coefficients of the interacting electron system. Occupation numbers
and nearest-neighbor overlap coefficients are taken into account. Both
the potential and the density are thus characterized by two parameters
per site. Considering the
complete Hamiltonian (\ref{Hamiltonian1D}), we can obtain an initial
guess by setting $U=0$. Table \ref{E-of-U} indicates that the
corresponding densities are relatively close and we can
therefore safely assume this to be a good starting point. The
eigenstates of the one-particle Hamiltonian $\hat{H}_s$ are
calculated by exact diagonalization. We have checked that the
optimization is stable and insensitive to any reasonable choice of
the initial guess.

We judge the quality of an approximation to the true density by
summing the squared deviations between the approximate and the
true density coefficients. Only the $M$ independent coefficients are
considered; the rest is determined by
the intrinsic symmetry $n_{ij} = n_{ji}$ and the
spatial symmetry with respect the center of the chain. As the hopping
parameters satisfy the same symmetry conditions, only $M$ are in fact
independent.

The results of the optimization for the half-filled eight-site chain
with $t=1$ and $U=4$ are shown in Table \ref{optdensM8}, where we
give the respective on-site energies and hopping parameters
and compare the density of the best-fit Kohn-Sham system
with the true density of the interacting system and that of the
initial trial Hamiltonian with $t=1$ and $U=0$. Although the
Kohn-Sham system gives the correct occupation numbers $n_i$, it is
evidently {\em unable to reproduce the correct overlap
coefficients\/} and thereby simulate the localization effect: the
density is {\em not\/} noninteracting {\it v} representable.
In fact, the overall quality of the approximation to the density, given
by the sum of the deviation squares {\it SUMSQ}, is only little better
than the approximation that is obtained by simply neglecting the
interaction in the first place and setting $U=0$.

\begin{table}
\caption{Hopping parameters, scaled so that $t_{45}=-1$, and density
of the Kohn-Sham system (KS) compared with the original system with
($U=4$) and without ($U=0$) interaction for the eight-site Hubbard chain.
The nonzero deviation between the overlap coefficients corresponding to
$U=4$ and KS, also represented by {\it SUMSQ}, indicates that the density
of the interacting electron system is {\em not\/} noninteracting {\it v}
representable.} \label{optdensM8}
\begin{tabular}{lrrr}
& $U=4$ & $\;\;$KS$\;\;$ & $U=0$ \\
\hline
$e_1 = e_8$ & 0 & 0 & 0 \\
$e_2 = e_7$ & 0 & 0 & 0 \\
$e_3 = e_6$ & 0 & 0 & 0 \\
$e_4 = e_5$ & 0 & 0 & 0 \\
$t_{12} = t_{78}$ & -1 & -0.027 & -1 \\
$t_{23} = t_{67}$ & -1 & -0.255 & -1 \\
$t_{34} = t_{56}$ & -1 & -1.021 & -1 \\
$t_{45}$ & -1 & -1.000 & -1 \\
\\
$n_1 = n_8$ & 1 & 1 & 1 \\
$n_2 = n_7$ & 1 & 1 & 1 \\
$n_3 = n_6$ & 1 & 1 & 1 \\
$n_4 = n_5$ & 1 & 1 & 1 \\
$n_{12} = n_{78}$ & 0.641 & 0.648 & 0.862 \\
$n_{23} = n_{67}$ & 0.370 & 0.428 & 0.495 \\
$n_{34} = n_{56}$ & 0.589 & 0.819 & 0.758 \\
$n_{45}$ & 0.391 & 0.503 & 0.529 \\
\\
{\it SUMSQ} & & 0.069 & 0.112 \\
\end{tabular}
\end{table}

It is interesting to note that the Kohn-Sham system approaches the
true density best at the ends of the chain, whereas towards the
center, it is little different from the density of the $U=0$ system.
This fact is reflected by the effective hopping parameters: towards
the ends of the chain, they drop substantially to
approach the atomic limit, which effectively reduces the overlap. It
is this contribution that is responsible for the decrease of {\it
SUMSQ} in the first place. In the center of the chain, however, the
hopping parameters are almost constant and the density resembles
that of the $U=0$ system.

\begin{figure}
\epsfxsize=8.1cm \centerline{\epsfbox{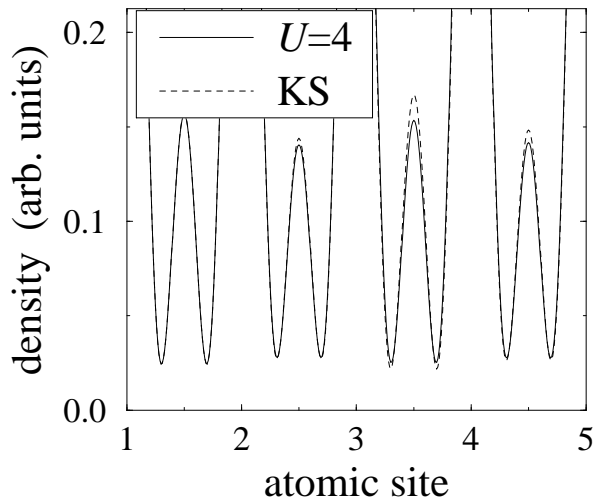}}
\bigskip
\caption{Electron density of the eight-site Hubbard chain ($U=4$)
together with that of the Kohn-Sham system that reproduces it as
closely as possible (KS), visualized by assuming a particular
form of orbitals (see the text). There is good agreement at the ends of
the chain (site 1), but increasing deviation towards the center (site
5). The deviation illustrates the fact that the density of the
interacting system is {\em not\/} noninteracting {\it v}
representable.} \label{densU4KS}
\end{figure}

In Fig.\ \ref{densU4KS} we compare the real-space density of the
Kohn-Sham system with that of the interacting system. The
increasing deviation towards the center of the chain is clearly
visible. We have used real symmetric orbitals based on trigonometric
functions for the purpose of
this visualization only. These assume a maximum at the
position of the atom, have one zero at either side, and extend to the
nearest-neighbor atom. The main numerical work is, of course,
independent of a specific choice of orbitals.

Although we are using a nonlinear optimization routine, the chances
of finding a solution will depend on the ratio of independent
variables to residual functions. Multiplying the Hamiltonian by
an arbitrary factor or adding a constant to the on-site energies
only affects the eigenvalues but leaves the eigenstates and thereby
the density coefficients unchanged. For this reason, the overlap
throughout the chain cannot be lowered by generally choosing smaller
hopping parameters. Also, because of these two degrees of freedom in
the hopping matrix, the number of relevant variables is in fact just
$M-2$ instead of $M$. As for the target density coefficients, one
is trivially determined by the requirement that the occupation numbers
$n_i$ add up to $N$. Besides, the density must be constructable from the
wave function of $N$ interacting electrons, but this imposes complicated
nonlinear relations that can be expected to be of no relevance for the
optimization procedure. The number of effectively independent residual
functions is therefore reduced by just one to $M-1$ and so exceeds the
number of free variables. From a purely mathematical point of view, the
nonexistence of an effective potential is thus not surprising.

We have performed calculations for different chain lengths and always
obtained similar results. In conclusion, we note that in the case of
small $M$, the density of the one-dimensional Hubbard chain is {\em
not\/} noninteracting {\it v} representable because the Kohn-Sham
system is unable to simulate the increased localization of the
electrons. This is a systematic feature and will hold for
all nonzero $U$.

\subsection{The limit $M \to \infty$}

The infinite chain is translationally invariant. The Hamiltonian of
the interacting electron system must therefore be of the form
\begin{eqnarray}
\hat{H}
&=& -t \sum_{i=-\infty}^\infty \sum_\sigma
\left( c^\dagger_{i\sigma} c_{(i+1)\sigma}
+ c^\dagger_{(i+1)\sigma} c_{i\sigma} \right) \nonumber \\
&+& U \sum_{i=-\infty}^\infty \hat{n}_{i\uparrow}
\hat{n}_{i\downarrow} .
\end{eqnarray}
Without performing numerical calculations, we can conclude from our
previous studies that the interaction will lead to a certain reduction
of the overlap and to an increased localization of the electrons. In any
case, the density of the interacting system will be different from the
density that we obtain after dropping the interaction term in the
Hamiltonian.

As for the noninteracting Kohn-Sham system, the translational
invariance requires that the effective on-site energies and hopping
parameters are constant throughout the chain. This leaves just two
variables, a uniform $e^{\rm eff}$ and $t^{\rm eff}$. However, these are
related to the two trivial transformations described above that leave
the eigenstates unchanged; their values do not affect the density and may
be scaled to zero and $-t$. As a consequence, the electron density of the
Kohn-Sham system must necessarily be identical to that of the $U=0$
system and it must then be different from the density of the original
system. Again, the Kohn-Sham system proves incapable of simulating the
reduction of the overlap.

\subsection{Discussion}

So far, we have proved that for small $M$ as well as in the limit $M
\to \infty$, the density of the interacting Hubbard chain is {\em
not\/} noninteracting {\it v} representable. Although translational
invariance is not strictly satisfied in the case of a finite chain
length, from a physical point of view, the situation of an atom
far from the ends, where surface effects have little influence, is
no different from that of an atom in an infinite chain.
It is thus not surprising that the Kohn-Sham system should
approach the $U=0$ system in the central region of the
chain. In fact, this effect is already visible in the case
of $M=8$ and it should become even more dominant for greater $M$.

The argument concerning the limit $M \to \infty$ does not rely on the
value of $U$ and holds for any nonzero interaction. We have thus
conclusively shown that the density of the interacting Hubbard chain is
{\em not\/} noninteracting {\it v} representable for all values of $M$
and $U$. Furthermore, the same argument can be used for regular lattices
with nearest-neighbor hopping in dimensions other than one.

\subsection{A semiconductor model}

Another generalized model ``with parameters more appropriate for a
semiconductor'' has been studied in Ref.\ \onlinecite{GS-band-gap}. The
eigenvalue gap obtained by density-functional theory that reproduced the
occupation numbers correctly was shown to be in good agreement with the
true band gap, whereas the value obtained by a LDA within that framework
was much smaller. It was therefore concluded that the LDA rather than
the discontinuity in the exchange-correlation potential seemed to be
responsible for the numerical deviation found in DFT calculations, but
doubts have been raised concerning the implications for the original
formulation of density-functional theory.\cite{comment}

Mathematically, the semiconductor model is equivalent to a
Hubbard Hamiltonian in the presence of a spin-dependent potential and
with the possibility of spin flipping. It features two nondegenerate
orbitals, each of which forms its own energy band. The electrons can
move within these bands as well as change from one to the other. The
model is understood to represent a semiconductor without magnetic
properties. Therefore, the electrons are regarded as spinless
fermions and the energy levels are interpreted as a low-lying {\it
s} band and a {\it p} band. All orbitals are considered localized, so
that overlap is negligible except on the same site and between
nearest neighbors. The two levels are labeled by $\alpha, \beta \in
\{s,p\}$. For a finite one-dimensional chain, the
Hamiltonian is given by
\begin{eqnarray}
\hat{H}
&=& \sum_{i=1}^M \sum_\alpha e_{i\alpha} \hat{n}_{i\alpha}
+ \sum_{i=1}^M t_{is,ip} \left(
c^\dagger_{is} c_{ip} + c^\dagger_{ip}
c_{is} \right) \nonumber \\
&+& \sum_{i=1}^{M-1} \sum_{\alpha,\beta}
t_{i\alpha,(i+1)\beta} \left( c^\dagger_{i\alpha}
c_{(i+1)\beta} + c^\dagger_{(i+1)\beta} c_{i\alpha}
\right) \nonumber \\
&+& \sum_{\alpha,\beta} t_{M\alpha,1\beta} \left(
c^\dagger_{M\alpha} c_{1\beta} + c^\dagger_{1\beta}
c_{M\alpha} \right) + U \sum_{i=1}^M \hat{n}_{is}
\hat{n}_{ip} .
\end{eqnarray}
The first term contains the on-site energies $e_{i\alpha}$,
the second and third govern the hopping between orbitals at the same
site and between nearest neighbors. Likewise, the following term has
been introduced to connect the end sites. As the orbitals are no
longer degenerate, all hopping parameters must depend both on spatial
position and the energy bands involved. The last contribution to the
Hamiltonian is the on-site Coulomb interaction.

The density-functional formalism derived for the Hubbard Hamiltonian
can be applied with minor changes, which account for the fact
that the two orbitals differ. In particular, we have to
consider occupation numbers and overlap coefficients for each pair of
orbitals separately:
\begin{equation}
n_{i\alpha,j\beta} = \langle \Psi |
c^\dagger_{i\alpha} c_{j\beta} | \Psi \rangle \quad
\mbox{or} \quad n_{i\alpha,j\beta} = \sum_{\gamma=1}^N
\langle \psi_\gamma | c^\dagger_{i\alpha} c_{j\beta} |
\psi_\gamma \rangle .
\end{equation}

We take parameters for the interacting system from Ref.\
\onlinecite{GS-band-gap}. The energy levels are $e_{is} = -4.0$ and
$e_{ip} = 0.0$. At the end points, the {\it s} state is raised by $1.0$
and the {\it p} state is lowered by $0.4$ to simulate surface effects.
There is no hopping between bands on the same site; the sign and the
magnitude of the other hopping parameters are consistent with a picture
of real symmetric {\it s} and antisymmetric {\it p} orbitals and set to
$t_{is,(i+1)s} = -1.8$, $t_{ip,(i+1)p} = 1.0$, and $t_{is,(i+1)p} =
-t_{ip,(i+1)s} = 1.2$. The end points are joined only by $t_{Ms,1p} =
-t_{Mp,1s} = 1.2$. Finally, the interaction parameter is $U = 4.0$.

The symmetries of the chain are more subtle than in the Hubbard case.
When varying the effective parameters in the one-particle Hamiltonian
$\hat{H}_s$, we constrain $e^{\rm eff}_{i\alpha} =
e^{\rm eff}_{(M+1-i)\alpha}$. Due to the different signs of the orbital
lobes, the hopping parameters are related by $t^{\rm eff}_{is,
ip} = -t^{\rm eff}_{(M+1-i)s, (M+1-i)p}$ and
$t^{\rm eff}_{i\alpha,(i+1)\beta} = \mp t^{\rm eff}_{(M-i)\beta,
(M-i+1)\alpha}$ with $-$ for $\alpha \neq \beta$ and $+$ otherwise.
Likewise, at the end points, $t^{\rm eff}_{M\alpha, 1\beta} =
-t^{\rm eff}_{M\beta, 1\alpha}$ for $\alpha \neq \beta$. The density
coefficients are related by the very same symmetries. A measure for
the accuracy of an approximation to the density is again given by
the sum of squares {\it SUMSQ} taken over the deviations of all
independent coefficients.

Calculated values of {\it SUMSQ} divided by the chain length $M$ are
listed in Table \ref{accuracy}. The columns correspond to (i) the
best-fit Kohn-Sham system, (ii) the system studied in Ref.\
\onlinecite{GS-band-gap}, in which merely the on-site energies are
varied so as to reproduce the occupation numbers of the original
system exactly, and (iii) the noninteracting system with the same
hopping parameters as the original interacting chain, which is used
as the starting point for the numerical optimization.

\begin{table}
\caption{Deviation between the independent density
coefficients of the nine-site semiconductor chain and those
of the Kohn-Sham system (KS), the noninteracting
system with the same occupation numbers ($n_{i\alpha}$
exact), and the original system with interaction set to zero
($U=0$), expressed through {\it SUMSQ} per site.
Although the approximation becomes better with increasing
$M$, the density of the interacting system is {\em not\/}
noninteracting {\it v} representable.} \label{accuracy}
\begin{tabular}{cccc}
$M$ & KS & $n_{i\alpha}$ exact & $U=0$ \\
\hline
3 & $0.477 \times 10^{-4}$ & $2.331 \times 10^{-4}$ & 0.0379 \\
5 & $0.118 \times 10^{-4}$ & $1.335 \times 10^{-4}$ & 0.0616 \\
7 & $0.049 \times 10^{-4}$ & $1.006 \times 10^{-4}$ & 0.0759 \\
9 & $0.034 \times 10^{-4}$ & $0.812 \times 10^{-4}$ & 0.0826 \\
\end{tabular}
\end{table}

Although {\it SUMSQ} assumes significantly lower values than it did
for the Hubbard model, despite the seven-fold increase in density
coefficients, it is still far from zero. For the chain lengths
studied here, the density of this interacting semiconductor model
is therefore {\em not\/} noninteracting {\it v} representable. To
give an impression of the numerical deviations, Table
\ref{central-site} lists the density coefficients for the central
site of the nine-site chain.

It seems that the approximation becomes better for increasing $M$,
but the data are not sufficient for a reliable extrapolation. A formal
repetition of our earlier argument regarding the limit $M \to \infty$
fails because translationally invariant systems will still retain
five relevant hopping parameters that determine the density. On the
other hand, particle conservation only helps to reduce the number of
effectively independent density coefficients to six, so it seems
probable that even in this limit the density is {\em not\/}
noninteracting {\it v} representable. We have performed calculations
for closed, translationally invariant ring chains up to $M=9$ that are
governed by the same ratio of significant hopping parameters and density
coefficients. None of these systems was found to be noninteracting
{\it v} representable.

\begin{table}
\caption{Density coefficients for the central site in a nine-site
semiconductor chain. The columns correspond to the true
interacting system ($U=4$), the Kohn-Sham system
(KS), the noninteracting system with the correct occupation
numbers ($n_{i\alpha}$ exact), and the initial-guess system with
interaction set to zero ($U=0$).} \label{central-site}
\begin{tabular}{lrrrr}
& $U=4\,$ & KS$\,\;\;$ & $n_{i\alpha}$ exact $\!\!\!$ & $U=0\,$ \\
\hline
$n_{5s}$ & 0.9209 & 0.9215 & 0.9209 & 0.7021 \\
$n_{5p}$ & 0.0782 & 0.0780 & 0.0782 & 0.2998 \\
$n_{5s,5p}$ & 0.0000 & 0.0000 & 0.0000 & 0.0000 \\
$n_{5s,6s}$ & 0.0360 & 0.0367 & 0.0363 & 0.1990 \\
$n_{5p,6p}$ & -0.0347 & -0.0352 & -0.0353 & -0.2010 \\
$n_{5s,6p}$ & 0.1641 & 0.1647 & 0.1671 & 0.2223 \\
$n_{5p,6s}$ & -0.1636 & -0.1640 & -0.1668 & -0.2214 \\
\end{tabular}
\end{table}

To double-check these findings, we have also taken a completely
different approach by varying the eigenvectors of the Hamiltonian
matrix directly and without further constraints. A Hamiltonian
constructed in this way corresponds to the unphysical case of hopping
not being restricted by distance. But although the number of variable
hopping parameters grows exponentially with $M$, even the overlap
between nearest neighbors cannot be reproduced accurately for $M
\leq 7$. The formal inclusion of next-nearest-neighbor hopping in
the Kohn-Sham system is therefore no solution.

\section{Conclusion}

We have reexamined the treatment of strongly correlated
Hubbard-type models within density-functional theory. A formulation
of the basic theorems that is related as closely as possible to the
conventional scheme through second quantization of the external
potential and the density has been provided.
Numerical calculations and qualitative arguments have shown that the
electron density of the original Hubbard model with nonzero
interaction is {\em not\/} noninteracting {\it v} representable. The
same result was found for a related semiconductor model that has
been used in a study of the band-gap problem.

\acknowledgments

We wish to thank P.~H.~Dederichs, O.~Gunnarsson, and K.~Sch\"onhammer
for helpful comments.
This work was supported by The Royal Society, the Science and
Engineering Council, and the European Community program Human
Capital and Mobility through Contract No.\ CHRX-CT93-0337. One of us
(A.S.) wishes to thank the Studienstiftung des Deutschen Volkes and
the Deutscher Akademischer Austauschdienst for financial support.

\end{document}